\documentclass[12pt]{article}

\usepackage{setspace}
\usepackage[colorlinks=true, linkcolor=blue, urlcolor=blue, citecolor=blue]{hyperref}
\usepackage{booktabs}
\usepackage[round,authoryear]{natbib}
\usepackage{listings}
\usepackage{geometry}
\usepackage{pgfplots}
\pgfplotsset{compat=1.18}
\usepackage{float}
\usepackage{amssymb, amsmath, mathtools}
\usepackage{csquotes}
\usepackage{subfig}
\usepackage{eurosym}
\usepackage{enumitem}
\usepackage{siunitx}
\usepackage{graphicx}    
\usepackage{placeins}   
\usepackage{adjustbox}  
\usepackage{booktabs}   
\usepackage{caption}    
\usepackage{tabularx}
\usepackage{array}
 
\usepackage{longtable} 
\usepackage{capt-of} 

\usepackage{fancyhdr}
\usepackage[normalem]{ulem}

\usepackage{booktabs}
\usepackage{wasysym} 

\title{Decision and Gender Biases in Large Language Models: A Behavioral–Economic Perspective}

\author{Luca\ Corazzini\\DEMS, University of Bicocca
\and Elisa\ Deriu\\DEMS, University of Bicocca 
\and Marco\ Guerzoni\thanks{Corresponding author: Marco Guerzoni, DEMS, University of Bicocca, Piazza dell'Ateneo Nuovo 1, 20126, Milan, Italy. Email: \href{mailto:marco.guerzoni@unimib.it}{marco.guerzoni@unimib.it}.}\\DEMS, University of Bicocca and BETA, University of Strasbourg}

\date{}

\begin{document}
\maketitle

\begin{abstract}
Large language models (LLMs) increasingly mediate economic and organisational processes, from automated customer support and recruitment to investment advice and policy analysis. These systems are often assumed to embody rational decision-making free from human error; yet they are trained on human language corpora that may embed cognitive and social biases. This study investigates whether advanced LLMs behave as rational agents or whether they reproduce human behavioural tendencies when faced with classic decision problems. Using two canonical experiments in behavioural economics, the ultimatum game and a gambling game, we elicit decisions from two state-of-the-art models, \textit{Google~Gemma--7B} and \textit{Qwen}, under neutral and gender-conditioned prompts. We estimate parameters of inequity aversion and loss aversion and compare them with human benchmarks. The models display attenuated but persistent deviations from rationality, including moderate fairness concerns, mild loss aversion, and subtle gender-conditioned differences. 
\end{abstract}


\section{Introduction}
Large language models (LLMs) are reshaping economic and organisational domains through their human-like abilities, increasingly taking roles once held by humans in areas such as customer service, education, and finance \citep{mollick2024co}. This raises a key question: do LLMs act as rational economic agents, or do they internalise human cognitive and social biases?

While algorithms are expected to be rational and bias-free, LLMs are trained on vast, uncurated corpora containing stereotypes and distortions \citep{rastogi2022deciding}. As a result, they may reproduce such biases at scale. Behavioural economics provides a framework to study these deviations through experiments on fairness, risk, and gender differences \citep{FehrSchmidt1999,KahnemanTversky1979,EckelGrossman2008}.

We extend this framework to artificial agents, asking whether LLMs display human-like behavioural patterns and whether these depend on contextual cues such as gender. Using the \emph{ultimatum} and \emph{gambling} games with two models, \textit{Google-Gemma-7B} (Gemma) and \textit{Qwen-2.5-32B-Instruct-AWQ} (Qwen) we estimate parameters of inequity aversion and prospect theory. LLMs occupy an intermediate position between rational and human-like behaviour: they are less biased than people yet deviate from expected-utility norms. This study aligns with \citet{ross2024examining}, who found consistent behavioural patterns across models but did not investigate gender effects. Regarding gender bias, our work is conceptually related to \citet{wan2023kelly}, though the latter lies outside an experimental framework. Section~\ref{sec:methods} outlines methods, Section~\ref{sec:results} presents findings, and Section~\ref{sec:discussion} concludes

\section{Methodology}\label{sec:methods}

Large Language Models (LLMs) are probabilistic systems that generate text by predicting the most likely continuation of a given input sequence. Their behavior is determined by the prompt, which specifies the context and task to be performed. The temperature parameter regulates the stochasticity of the model’s output: lower values lead to more deterministic responses, while higher values introduce variability and diversity across repetitions. In all experiments, prompts were carefully designed to ensure comparability across conditions, and each configuration was replicated with a temperature of 1 to allow for controlled variation in responses.

We evaluate two open-source LLMs which were queried via publicly available endpoints without further fine-tuning.Each model was prompted to play one of the games described below under three conditions. The first follows the standard formulation of the game, serving as a baseline. In the other two conditions, the LLM is instructed to assume the role of either a male or a female player. Specifically, we replicated with two gendered identities — male (Joseph, \textit{he/him}) and female (Kelly, \textit{she/her}) — to test whether, under identical conditions, systematic differences would emerge in inequality aversion, risk propensity, probability weighting, and loss aversion. 

We employ two canonical behavioral–economic tasks to study fairness and risk preferences: the Ultimatum Game \citep{GuthSchmittbergerSchwarze1982} played by \emph{Gemma} and a Gambling Game played by \emph{Qwen}, both described in the following subsections. Technical details on model specifications and implementation including drivers of model choices are provided in the Supplementary Materials.

\subsection{The Ultimatum Game (UG)}
UG models bilateral bargaining between a proposer and a responder. In the present implementation, UG was conducted with varying stake sizes, with pool values ranging from 2 to 10. For each pool, the proposer selected an integer offer within the available amount, while the responder decided whether to accept or reject it. Each configuration was repeated 100 times at a model temperature of 1 to ensure sufficient variability and robustness in the generated responses.

The analysis considers three aspects: the mean offer as a proportion of the total stake; the consistency of offers within a given stake (intra–pool consistency); and the consistency of offers across different stakes (inter–pool consistency). 
Following \citet{FehrSchmidt1999}, we estimate the parameters according to:
\begin{equation}
U_i(x_i,x_j) = \begin{cases}
 x_i - \alpha_i\,(x_j - x_i), & \text{if } x_i < x_j,\\
 x_i - \beta_i\,(x_i - x_j), & \text{if } x_i \geq x_j,
\end{cases}
\label{eq:fehrschmidt}
\end{equation}
where $x_i$ is the payoff to player~$i$, $\alpha_i$ measures aversion to disadvantageous inequity and $\beta_i$ aversion to advantageous inequity. 
Concerning the receiver, for each UG configuration, we compute the \emph{switching point}, namely the lowest offer for which the model accepts more than 50\% of the proposals she receives. This threshold is then used to estimate the parameter $\alpha$, aversion to inequity when disadvantage. The average offer is employed to estimate $\beta$, which measures aversion to advantageous inequity, also known as guilt parameter; the larger this parameter, the more willing the proposer is to offer substantial amounts.
Typical estimates obtained by analyzing human choices in experiments are $\alpha_i \approx 0.6$--$1.0$ and $\beta_i \approx 0.2$--$0.3$ \citep{FehrSchmidt1999}.

\subsection{The Gambling Game (GG)}
GG consists of a series of binary choices between risky and certain payoffs. The experiment is structured across three domains, defined by the nature of the lottery outcomes. In the gain domain, both outcomes are positive ($x > 0, y > 0$); in the loss domain, both are negative ($x < 0, y < 0$); and in the mixed domain, one outcome is positive and the other negative ($x > 0, y < 0$), allowing for the isolation of loss aversion through symmetric comparisons around the reference point. The magnitudes are set to $m \in \{20, 35, 50, 70, 100, 140, 200\}$. For each $m$, four probabilities are tested in the gain and loss domains and one in the mixed domain, for a total of 56 configurations. Each configuration is replicated with a temperature of 1 and 100 repetitions.
These choices are analyzed by using Prospect Theory \citep{KahnemanTversky1979}, which posits a value function that is concave for gains, convex for losses and kinked at the reference point. As subjective value function, we assume the functional form of Tversky and Kahneman's cumulative prospect theory\citep{TverskyKahneman1992}:
\begin{equation}
v(x) = \begin{cases}
 x^{\alpha}, & x \geq 0,\\
 -\lambda\,(-x)^{\beta}, & x < 0,
\end{cases}
\label{eq:prospect}
\end{equation}
and with $\pi(p_i)$ representing the subjectively weighted probability:
\begin{equation}
\pi(p_i) =
\begin{cases}
\pi^{+}(p_i) =
\dfrac{p_i^{\phi^{+}}}
{\bigl(p_i^{\phi^{+}} + (1 - p_i)^{\phi^{+}}\bigr)^{1/\phi^{+}}}, & x_i \geq 0,\\[10pt]
\pi^{-}(p_i) =
\dfrac{p_i^{\phi^{-}}}
{\bigl(p_i^{\phi^{-}} + (1 - p_i)^{\phi^{-}}\bigr)^{1/\phi^{-}}}, & x_i < 0.
\end{cases}
\label{eq:prospect_weight}
\end{equation}
Together, the value function, $v(x)$, and the weighted function, $\pi(p_i)$, define the following utility function:
\begin{equation}
U = \sum_{i} \pi(p_i) \, v(x_i),
\label{eq:prospect_general}
\end{equation}

\noindent
where $\alpha$ and $\beta$ capture the curvature of the value function (risk attitude for gains and losses),
$\lambda$ measures loss aversion, and $\phi^{+}$, $\phi^{-}$ determine the distortion of favorable and unfavorable probabilities, respectively.

We estimated the parameters $(\alpha,\beta,\lambda, \phi^{+}, \phi^{-})$ by fitting the models' acceptance frequencies to the value function in equation \eqref{eq:prospect}. The certainty equivalent  of each lottery was derived from the point at which the model expressed indifference between the gamble and the certain payoff. We then used nonlinear least squares to find the parameters that minimised the squared deviation between predicted and observed certainty equivalents, separately for gains and losses and for each framing condition. Cases where the model produced non-numeric or ambiguous outputs were excluded (less than 2\% of responses). 
 
Benchmark values associated with the choices of human participants in experiments are $\lambda \approx 2.25$ and $\alpha \approx \beta \approx 0.88$ \citep{TverskyKahneman1992}.

\section{Results}\label{sec:results}

\subsection{UG}
The analysis of \emph{Gemma}’s behaviour reveals patterns consistent with both \citet{ross2024examining} and experimental evidence with human subjects \citep{FehrSchmidt1999}. Three aggregate metrics are considered: the acceptance threshold $s$, the inequity‐aversion parameter $\alpha$ and the guilt parameter $\beta$. For \emph{Gemma} the acceptance threshold is close to 1, which means that the responder accepts any positive offer. This leads to an estimated $\alpha \approx 0.454$, which is substantially lower than the values reported for humans. Such a low estimate reflects a weak inclination to punish unfavorable splits. Conversely, the estimated $\beta \approx 0.542$ is relatively high, indicating a strong aversion to retaining an unfair advantage: the model shows a bias toward guilt, preferring to share more generously even though it is less willing to reject unfair offers. Overall, \emph{Gemma} thus behaves more like a utilitarian agent—accepting almost all positive offers, while still signaling a concern for equitable outcomes.
An interesting result is the inversion of the typical human pattern: while humans generally display $\alpha > \beta$, indicating stronger aversion to disadvantageous than to advantageous inequality, LLMs show the opposite ($\beta > \alpha$). This suggests that LLMs place more emphasis on avoiding excessive advantage than on rejecting unfair offers, exhibiting a form of fairness biased toward the other’s benefit and closer to the \textit{homo economicus}' efficiency-oriented behavior. Overall, this results in a model less inclined to perceive disadvantageous offers as unacceptable, yet more sensitive to avoiding situations of excessive gain, thus revealing — albeit in its own way — a stronger aversion to inequality compared to human subjects.

\subsubsection{Gender-Differentiated UG}
Gender conditioning yielded systematic but small shifts. The gender-differentiated UG shows that \emph{Gemma} behaves consistently across all proposer–responder pairings, with average offers  between 46\% and 48\%, close to an equal split. Gender mainly produces a level shift rather than a structural change in offer dynamics. 

Variability measures indicate slightly more stable behavior when the responder is male ($E[\sigma_N]=5.95\%$) and greater dispersion when the responder is female ($E[\sigma_N]=7.74\%$), suggesting that male recipients elicit more consistent strategies. Acceptance thresholds are marginally higher for male responders ($s^*\approx1.52$) than for female responders ($s^*\approx1.46$).

Across all gender configurations, the inequality aversion pattern $\beta > \alpha$ persists, confirming a bias toward avoiding excessive advantage rather than punishing unfairness. Female proposers show slightly lower $\alpha$ and higher $\beta$ values, indicating a more “other-regarding” orientation, whereas male proposers display higher $\alpha$, reflecting mild resistance to disadvantageous outcomes. Overall, gender cues modulate but do not overturn the model’s structure: \textit{Google Gemma-7B} remains closer to the \textit{homo economicus}' efficiency-oriented behavior than to the human benchmark $\alpha > \beta$.

\subsection{GG}

\textit{Qwen} was propted to play the game with the following results. In the gain domain, identification mainly concerns the curvature of the value function and the weighting of favorable probabilities $(\alpha, \phi^{+})$; in the loss domain, it involves the curvature in losses and the weighting of unfavorable probabilities $(\beta, \phi^{-})$; in the mixed domain, all parameters $(\alpha, \beta, \phi^{+}, \phi^{-}, \lambda)$ jointly contribute, as gains and losses are combined with distinct probability weights and asymmetric loss aversion $\lambda$.

\begin{table}
\centering
\caption{Point estimates in GG}
\label{tab:gg_params_point_clean}
\begin{tabular}{lccccc}
\hline
\textbf{Domain} & \(\alpha\) & \(\beta\) & \(\phi^{+}\) & \(\phi^{-}\) & \(\lambda\) \\
\hline
GAIN     & 1.062 & --    & 1.001 & --    & -- \\
LOSS+MIX & --    & 0.932 & --    & 0.800 & 1.542 \\
\hline
\end{tabular}
\end{table}
\noindent

Table~\ref{tab:gg_params_point_clean} reports the point estimates by domain. Parameters not listed are not relevant in that domain by model definition. Overall, the model displays near-risk-neutral behavior for gains ($\alpha=1.062$) and nearly linear probability weighting ($\phi^{+}=1.001$), aligning closely with the \textit{homo economicus} benchmark. For losses, curvature and weighting remain moderate ($\beta=0.932$, $\phi^{-}=0.800$), while loss aversion is present but attenuated ($\lambda=1.542$), requiring a gain about 1.5 times larger than the loss to break even. Compared with typical human estimates ($\alpha,\beta\approx0.88$; $\lambda\approx2.25$), the model exhibits weaker prudence and lower loss sensitivity, forming a hybrid profile: rational and linear on the gain side, yet mildly loss-averse and probability-distorting on the loss side.

\subsubsection{Gender-Differentiated GG}

\begin{table}[H]
\centering
\caption{Point estimates and model fit by gender in GG}
\label{tab:gg_gender_params_summary}
\begin{tabular}{lccccccc}
\hline
\textbf{Gender} & \(\alpha\) & \(\phi^+\) & \(\lambda\) & \(\beta\) & \(\phi^-\) & \(R^2(\alpha,\phi^+)\) & \(R^2(\lambda,\beta,\phi^-)\) \\
\hline
male   & 1.091 & 0.969 & 0.800 & 0.600 & 0.700 & 0.869 & 0.608 \\
female & 1.091 & 0.970 & 0.800 & 0.600 & 0.700 & 0.870 & 0.608 \\
\hline
\end{tabular}
\end{table}

Table~\ref{tab:gg_gender_params_summary} reports estimated parameters when differentiating by gender. Specifically, the columns of the table report the curvature parameter (\(\alpha\)), the probability weight on gains (\(\phi^+\)), the loss‐aversion coefficient (\(\lambda\)), the curvature in losses (\(\beta\)), the probability weight on losses (\(\phi^-\)), and the \(R^2\) fit statistics for gains and for the combined loss and mixed domains. The estimates reveal a striking overlap between male and female identities. In the gain domain, the value function is essentially linear (\(\alpha \approx 1\)) and probability weighting is nearly uniform (\(\phi^+ \approx 1\)), with a high model fit (\(R^2 \approx 0.87\)).  This suggests that the model evaluates favourable prospects almost at expected value and shows no appreciable gender difference.  In the loss and mixed domains the loss‐aversion coefficient (\(\lambda = 0.80\)) lies below unity, indicating that losses are weighted less than gains; the value function for losses (\(\beta = 0.60\)) is concave, which traditionally reflects a willingness to take risks to avoid a sure loss. However, the probability weighting for losses (\(\phi^- = 0.70\)) and the modest fit (\(R^2 \approx 0.61\)) show that behaviour in this domain is noisier and less consistent than in gains.

Contrary to some empirical findings that women exhibit greater risk and loss aversion \citep{croson2009,sarin2016,dawson2023}, and consistent with other studies reporting negligible gender effects \citep{bouchouicha2019}, the absence of divergence here supports the hypothesis that abstract, impersonal tasks like GG are insensitive to gender.  The gain domain remains well explained by the model, whereas losses and mixed scenarios are noisier; the near perfect overlap between male and female estimates strengthens the view that gender differences emerge primarily in social and interactive contexts (such as the Ultimatum Game) rather than in pure risk preferences. Overall, the gender analysis confirms that the model exhibits near risk neutrality in gains, modest loss aversion, and that gender labels do not influence its preferences in this abstract setting.

\section{Conclusion}\label{sec:discussion}

Our findings show that modern LLMs display structured behavioural patterns in economic decisions without explicit programming of fairness or risk preferences. These tendencies, inherited from human language data, suggest that LLMs internalise social heuristics: they penalise unfairness and exhibit loss aversion, though less strongly than humans. Their bounded rationality implies both limitations and potential: while not fully “rational,” they show milder biases that might yield more balanced outcomes. Gendered prompts reveal modest but significant stereotype effects. We also propose a replicable pipeline to estimate behavioural parameters in LLMs, offering a behavioural-economics perspective complementary to standard NLP benchmarks.

\section*{Declaration of Generative AI and AI-Assisted Technologies in the Manuscript Preparation Process}

During the preparation of this work the authors used Claude (Anthropic) in order to perform English revision and LaTeX formatting, as well as Python debugging. After using this tool, the authors reviewed and edited the content as needed and take full responsibility for the content of the published article.

\bibliographystyle{plainnat}   
\bibliography{reference}

\end{document}